# Classification analysis of transition-metal chalcogenides and oxides using quantum machine learning

*Kurudi V Vedavyasa and Ashok Kumar\**

Department of Physics, School of Basic Sciences, Central University of Punjab, Bathinda, 151401, India.

\*Corresponding author: ashokphy@cup.edu.in

(March 05, 2024)

**Abstract**

Quantum machine learning (QML) leverages the potential from machine learning to explore the subtle patterns in huge datasets of complex nature with quantum advantages. This exponentially reduces the time and resources necessary for computations. QML accelerates materials research with active screening of chemical space, identifying novel materials for practical applications and classifying structurally diverse materials given their measured properties. This study analyzes the performance of three efficient quantum machine learning algorithms viz., variational quantum eigen solver (VQE), quantum support vector machine (QSVM) and quantum neural networks (QNN) for the classification of transition metal chalcogenides and oxides (TMCs &TMOs). The analysis is performed on three datasets of different sizes containing 102, 192 and 350 materials with TMCs and TMOs labelled as +1 and -1 respectively. By employing feature selection, classical machine learning achieves 100% accuracy whereas QML achieves the highest performance of 99% and 98% for test and train data respectively on QSVC. This study establishes the competence of QML models in materials classification and explores the quantum circuits in terms of over-fitting using the circuit descriptors expressibility and entangling capability. In addition, the perspectives on QML in materials research with noisy intermediate scale quantum (NISQ) devices is given.

**Keywords:**

Variational quantum eigensolver, quantum support vector machines, quantum neural networks, parameterized quantum circuits, transition metal chalcogenides, transition-metal oxides**.**

## 1. Introduction

The past decade has seen a phenomenal shift in computational technologies with the advent of two key areas i.e. machine learning and quantum computing, being vehemently explored and put to use in different fields.[1-6] These techniques are developed to advance the computational abilities, drive user defined services and increase the efficiency and accuracy of the computers along with polynomial reduction in the time needed for processing data. Machine learning in particularly has the potential to analyze big datasets and identify subtle patterns underneath for developing a model, based on which future discoveries or properties can be predicted.[7] Quantum computing harnesses the power of quantum realm like annealing, entanglement and superposition to accelerate the computing by exponentially decreasing the resources as well as fostering accuracy, speed and security. Few benchmark studies have already proven the substantial advantage of quantum computing over classical ones for specific problem sets which is termed as 'Quantum speedup' or 'Quantum supremacy'.[8]

The combination of Machine Learning (ML) with Quantum computing is a new research discipline known as 'Quantum Machine Learning (QML)'. The aim of QML is to design such algorithms which are manifold times efficient when compared to their classical machine learning counterparts and apply them for the practical problems.[9] This has been established in theory and experiments as well, e.g., QML algorithms can search an unsorted database with N entries in time proportional to $\sqrt{N}$ whereas the exact classical counterpart takes time proportional to $N$ thus showing a quadratic speed-up in computations.[10] Similarly, using a QML algorithm called as "quantum basic linear algebra subroutines (QBLAS)" one can perform Fourier transforms over $N$ data points, invert $N \times N$ matrices and find their eigen values in time proportional to $log_2 N$. The best available classical algorithms take time proportional to $Nlog_2 N$ thus showing an exponential speedup in computations.[11] It has been verified that Quantum kernel estimation, an efficient QML algorithm, can accelerate supervised learning of the 67-featured cosmological supernova dataset to achieve 100% accuracy when performed on Google's Quantum processor 'Sycamore' for which the competitive classical classifiers substantially failed.[12] This approach was also tested in other fields of physics such as high energy physics,[13] solid-state physics[14] where potential results were seen using multiple-qubits. QML algorithms can notably become invaluable for data mining, data clustering, classification and regression.[15]

Also, the increase in the computational power has also advanced the field of material science to a great extent to include atomistic level of interactions and simulate a wide variety of materials in different lengths and time scales. But, the existing classical methods have reached their limiting potential in mimicking the complex nature of interactions and are proving to be computationally very costly.[16] The usage of machine learning techniques have fuelled the rapidly emerging applications in material science like atomistic simulations, materials imaging, spectral analysis and natural language processing of materials data.[17] It has also greatly improved the ability to identify the non-linear relationships among various material datasets.[18] The usage of QML algorithms in materials research can enhance the ability of simulating the actual chemical search space by including the quantum effects with greatly reducing the time for computations.[19] They can give an impetus to materials discovery[20] and play a vital role for calculating the electronic structures of various materials.[21-23] For ever increasing variety and complexity of the materials, QML techniques can greatly reduce the hardware and software problems by efficiently utilising the resources.[24]

In the past few decades, transition metal chalcogenides (TMCs) which encompass a large family of 2D materials have emerged as an important class of materials due to their potential significance in many technological applications. This includes optoelectronic applications,[25] photocatalytic applications,[26] biosensors,[27] energy storage devices,[28] active electrodes due to their high flexibility and mechanical strength[29] and waste water treatment including bacterial disinfection and organic contaminants degradation.[30] So, they occupy a very important role in materials research.

Transition Metal Chalcogenides are inorganic chemical compounds consisting of at least one chalcogen anion (O, S, Se, Te) and at least one more electropositive transition metal element (Groups 3 to12 excluding lanthanum and actinium series). Although all the group-16 elements are called chalcogens, the term TMCs is reserved for the transition metal compounds with sulfides, selenides and tellurides instead of oxides and polonium compounds since they extremely differ in their characteristics. TMCs have narrow bandgap or metallic characteristics, whereas transitional metal oxides (TMOs) are non-metallic, highly stable, abundantly available and have broader bandgap.[31] Thus, there is a need to classify the two classes given the interesting differences in many of their properties. The use of ML models to explore the multi-dimensional chemical space of these materials for obtaining the desired functionalities have been fruitful with very good accuracy. Many ML models are also being built for classification and regression purposes of TMCs and TMOs and their doped structures for predicting the phase stability, band gaps and conductivities.[32] ML has enormously improved the potential to probe these materials where trial and error experiments and high throughput calculations cannot comprehensively study the structures. Thus, it is also natural to utilize the quantum advantages offered by QML models to investigate the materials and enquire their performance and suitability in probing the properties of the increasing complexity in this field. Recently, it has been shown that the hybrid classical-quantum ML model can classify perovskite structures with a great accuracy thus establishing the ability of QML algorithms.[33]

The scope of present work is to test widely popular QML algorithms viz. Variational Quantum Eigen solver (VQE), Quantum Support Vector Algorithm (QSVA) and Quantum Neural Networks (QNN) for the classification of TMCs and TMOs. The input dataset is carefully constructed from the available materials in the open source Materials project database.[34]

## 2. Methodology

The workflow consists of first giving a brief description about the various QML algorithms and Parameterized Quantum Circuits (PQCs) that are used in this study. It is followed by 'Features selection' consisting of Random forest feature importance and Pearson correlation heatmap to select the most important features among the initial 10 features and discard the others for further study. Next, we test the performance on Classical Machine learning model and verify the results using K-fold cross validation to address over-fitting. Later, we employ different combinations of QML classifiers and feature maps to test the performance of each possible combination on test and train data of all the different sizes of datasets. Then we draw conclusions by comparing the results to identify the best performing scenario and analyse. Different components of the methodology are discussed below:

### 2.1. Quantum machine learning (QML) Algorithms

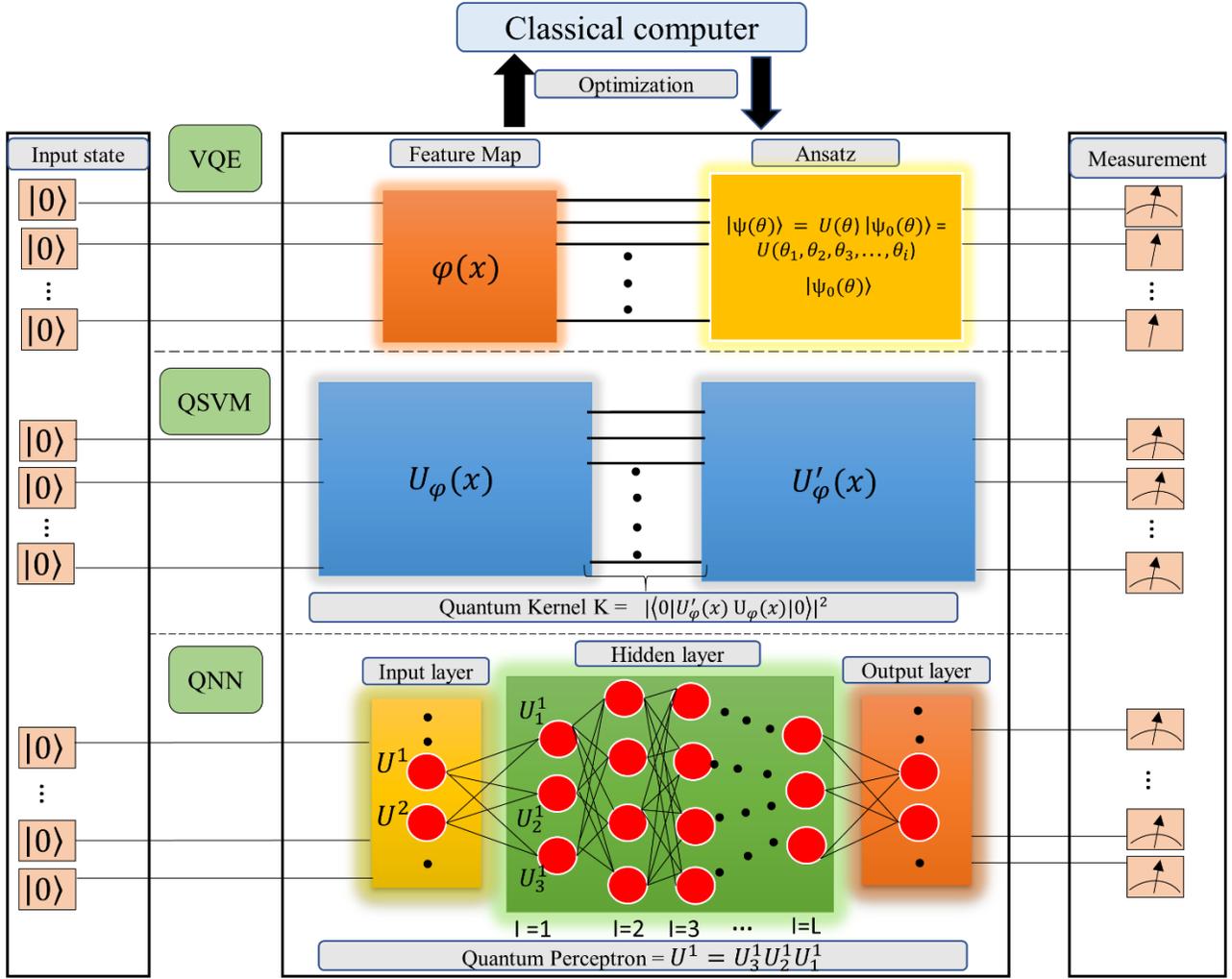

**Figure 1.** The block diagram of various quantum machine learning algorithms used in this study are shown along with their quantum circuits. VQE, QSVM and QNN stands for variational quantum eigen solver, quantum support vector machines and quantum neural networks respectively. The Quantum circuits are continuously optimized by updating their parameters from the classical optimizer working on a classical computer. The feature map encoding, the minimization of the Hamiltonian of an ansatz, quantum kernel estimation and quantum perceptron forward feed are performed on quantum computer.

There are a number of QML algorithms available in theory which are inspired from classical machine algorithms thus acting as their substitutes in quantum space and purely quantum mechanics based algorithms also exist.[35, 36] Any QML algorithm must be viable to realize on existing quantum computers or on Noisy Intermediate Scale Quantum (NISQ) devices that will be available in the near future.[37] The best way is to combine the quantum computers with classical computers for obtaining a hybrid quantum – classical model which seems to be the promising route towards achieving quantum advantage on NISQ devices.[38] The three algorithms which standout to be successful in this approach are variational quantum eigensolver (VQE), quantum kernel estimation (QKE) or quantum support vector machines (QSVM) and quantum neural networks (QNN).[39-41] Figure 1 shows the broad working of these algorithms. They leverage the classical resources efficiently coupled with quantum circuits to solve problems which are otherwise intractable on conventional computers.

VQE or variational quantum classifier (VQC) is a well-known hybrid quantum-classical algorithm in QML which is used in a number of quantum chemistry related problems.[42] It works on the basic quantum mechanics principle called Rayleigh-Ritz variational principle.[43] If $|\psi\rangle$ is the energy eigen state of any quantum mechanical system, then the Ground state energy $E_G$ is given by

$$E_G \geq \frac{\langle\psi|H|\psi\rangle}{\langle\psi|\psi\rangle} \qquad (1)$$

where H is Hamiltonian operator. Thus, we construct a quantum circuit where the input information can be effectively encoded into qubits in terms of parameters '$\theta$' (Parameterized Quantum circuit) whose combined state $|\psi(\theta_i)\rangle$ ($i$ is the number of parameters) acts as a trial wavefunction (ansatz). The combined state of qubits can be constructed using $e^{-i\theta\sigma}$ rotation gates where $\sigma$ is chosen from set of (X, Y, Z) Pauli gates.

$$|\psi(\theta_i)\rangle = U(\theta)|\psi_0(\theta)\rangle = U(\theta_1, \theta_2, \theta_3, \ldots, \theta_i)|\psi_0(\theta)\rangle \qquad (2)$$

where $U(\theta_i)$ represents different unitary quantum rotations or layers inside a given quantum circuit. The state $|\psi(\theta_i)\rangle$ needs to be minimized for obtaining the lowest energy ground state from which we can deduce the information for classifying the data into their respective classes.[44] The classical part comes into picture with the optimisation of the quantum state which requires a classical optimizer like COBYLA to continuously change the parameters by constructing a new quantum state in each iteration.[45]

QSVM or QKE is the quantum variant of the classical ML algorithm 'Support vector machines (SVM)'. Here, quantum kernels are constructed from the input data mapped into higher dimensional feature space. The mapping is done non-linearly onto the Hilbert space of the qubits. The quantum kernels are obtained by taking the unitary parameterized quantum circuit $U_\varphi(x)$ with the quantum state $|\psi(x)\rangle$ and performing the inner product with the inverted quantum circuit $U'_\varphi(x)$ with quantum state $|\psi'(x)\rangle$.[46]

Quantum Kernel 'K' = $|\langle\psi'(x)|\psi(x)\rangle| = |\langle 0|U'_\varphi(x)\,U_\varphi(x)|0\rangle|^2$ \qquad (3)

Later, K is evaluated by SVM which constructs the hyperplane in feature space to segregate the components.[47] The hyperplane thus constructed has the largest distance to the nearest training entries for any class. The same hyperplane used for classification of the training data is utilised to classify the kernel entries of the test dataset as well.

The kernel entries are evaluated on quantum computer whereas the construction of hyperplane from the support vector machines and its optimisation is done on classical computer. The performance of the model depends on the ability of hyperplane to classify kernel entries into their respective classes.

QNN draws its inspiration from classical neural networks (CNN). The complexity in training these CNNs in big data applications[48] enables us to incorporate the quantum advantages for restructuring the training process and developing the quantum versions.

The basic structure of a neural network consists of 3 components called input component, hidden component and output component. Figure 1 describes the structure of these components. All of them are constructed from 'artificial neurons' which are processed with a threshold function known as 'perceptron'. It helps in decision making with the pre-defined function value and generates the output. The collection of such artificial neurons results in a neural network. Thus, a neural network is built by one input layer with input vector $x \in \mathbb{R}^d$ and one output layer $y \in \mathbb{R}^m$ ('$m$' may or may not be equal to '$d$') with a number of in between hidden layers. The quantum neural network has 'quantum perceptron' that works through a series of arbitrary non-linear unitary quantum operators ($U$).[49]

Quantum perceptron $U^1 = U_3^1 U_2^1 U_1^1$ (4)

The cost function $C$ of QNN is defined as

$C = \frac{1}{N} \sum_{x=1}^{N} \langle \varphi_x^{out} | \rho_x^{out} | \varphi_x^{out} \rangle$ (5)

where $\varphi_x^{out}$ is the desired output $\rho_x^{out}$ is the obtained output from QNN. We need the help of a classical optimizer to minimize the above cost function by changing the parameters.

## 2.2. Parameterized Quantum circuits (PQCs)

'Parameterized Quantum circuits (PQC)'[50] can serve two purposes in QML:i) to encode the noramlized classical information into quantum hilbert space using 'quantum feature maps'[47] and ii) to provide the intial quantum state in the form of a quantum circut for training and optimisation of the QML model using 'ansatz'.[51]

In any ML model, for a set of data $\{x_i\}_i \in [N]$ where the samples $x_i \in \mathbb{R}^d$ for every $i$, a classical outcome which includes labels or classes $\{y_i\}_i \in [N]$ where $y_i \in \mathbb{R}^d$ is required. For a practical application of quantum computer one need to encode this classical data into quantum hilbert space. For this purpose, data encoding i.e. a non-linear feature transformation $\varphi$ using feature maps into a hilbert space $v$ i.e $\varphi: \mathbb{R}^n \to v$ has to be applied.[51]

In angle encoding method of dataencoding, the data is encoded into the angle of the qubits.[52] This is done by using pauli rotation gates (*I, X, Y, Z*) or their combinations like (*XY, YY, ZZ*) etc. in a quantum circuit along with entaglement among the qubits with the help of Hadamard gates *H*. The pauli feature map circuit transforms the input data $x \in \mathbb{R}^n$ where n is the feature dimension (number of features used) into:

$$U_{\varphi(x)} = exp[i \sum_{S \epsilon x} \varphi_s(\vec{x}) \prod_{i \epsilon S} P_i] \tag{6}$$

where $S$ is a set of qubit indices that describes connections in the feature map and $P_i \in (I, X, Y, Z)$. The parameters of data are mapped into quantum circuit as $\varphi_s$. It is given by

$$\varphi_s = \begin{cases} x_i & if\ S = (x_i) \\ \prod_{j \epsilon s}(\pi - x_j) & if\ |S| > 1 \end{cases} \tag{7}$$

The possible connections can be set using the entanglement and Pauli arguments available in the Qiskit library.[53] The widely used encoding for the real world quantum chemistry and material probems is 'ZZ-Feature encoding.' Any feature map can be generated for n qubits by applying the following unitary transformation:

$$u_\varphi(x) = U_\varphi(x) H^{\otimes n} U_\varphi(x) H^{\otimes n} \tag{8}$$

where $U_\varphi(x)$ is a unitary transformation already described above.

The choice of feature map influences the performance of the PQC. The quantification of the performance of a given PQC can be done in terms of two descriptors namely 'expressibility' and 'entangling capability'.[54] The power and capability of a particular PQC depends on its ability to uniformly reach the full Hilbert space. These descriptors guide us with the choice and design of PQCs and provide the inputs with potential limitations. Expressibility refers to the ability to explore the Hilbert space and entangling capability captures the non-trivial correlation in the quantum data. Expressibility can be quantified by comparing the fidelities of the true distribution corresponding to a PQC and the distribution obtained from the Haar random states. The true sample distribution can be constructed by repeatedly sampling the two sets of variational angles and then calculating their fidelity given by $F = \langle \psi_\theta | \psi_\varphi \rangle^2$. Haar random states[55] can be calculated analytically by

$$P_{Harr} = (N-1)(1-F)^{N-2} \tag{9}$$

where N is the dimension of the Hilbert space. Expressibility can be finally obtained by measuring the Kullback – Leibner divergence.[56] A smaller KL divergence value indicates the better ability to explore the Hilbert space. The entagling capability can be measured in multipe ways. One such method is Meyer-Wallach entanglement measure.[57]

The second PQC that we use here is 'ansatz'. An ansatz is a trial state wavefunction which acts like a starting point or an approximation for a classifier to optimize the predefined objective function. An ansatz is parameterized based on the weights specified for each entry.In variational classifiation, an ansatz acts like a variational circuit which on optimization yields the predicted label for the entry.[58] In quantum neural networks it can be considered as an equilvalent to classical neural networks consisting of multiple layers. The ansatz can also be considered as the subspace of a big Hilbert space which we look for a solution inside it.We can narrow down it to smaller and smaller subspaces by continuously adjusting the weights.Since our solution lies in the real space of the Hilbert space we discard the imaginary amplitudes and consider only real amplitudes of an ansatz.This can be specified in Qiskit by considering the 'Real amplitude ansatz'.

The next step in constructing the model after having feature maps and ansatz is to define an 'objective function' also known as 'Loss function'. It calculates the discrepancy between the obtained value after training and the predefined value.Thus, minimizing the objective function or minimizing the loss is achieved by continuously adjusting the parameters of the ansatz.In our classical- quantum model, this optimization process requires a classical optimizer which adjusts the parameters continuously.[59] This process of optimizing the circuit is known as 'training.'There are a variety of training methods available like gradient based methods, natural gradients, simultaneous perturbation stochastic approximation (SPSA) etc. In our study we used COBYLA optimizer which employs the linear approximation to handle the constraints and the objective function.[45] It is specifically useful for minimising the loss when the derivative of the objective function is unknown. For improving the model's performance we define a call back function which provides the Callback graph between loss and iteration.After a sufficient number of iterations of adjusting the parameters and training the circuit, the loss is minimized and the model is set to be trained.Then, we test our quantum-classical model for the test-data to caluculate the performance.We obtain the training and testing score seperately to compare the results from classical models. All the above analysis is executed on a classical system equipped with an Intel(R) Core(TM) i3-7020U CPU @ 2.30 GHz 4.0 GiB RAM running Python 3.11.5 with IBM Qiskit runtime, which included qiskit machine learning:0.7.2.

## 3. Results and Discussions:

We have constructed three different sizes of datasets with 102 materials, 192 materials and 350 materials consisting of TMCs and TMOs of the form $M_a X_b$ (a, b = 1,2,3,4; M = transition metal and X= O, S, Se, Te). For the purpose of supervised learning, TMCs are labelled as +1 and TMOs are labelled as -1. All the datasets have 10 initial features namely energy above convex hull (E-hull), energy of bandgap (E-band), energy of formation (E-form) including the structural properties of a single unit cell namely crystal parameter (a), volume (V), density (dens.) and M-X bond length (d[M-X]). Atomic properties of chalcogens namely atomic number A(X), atomic radius r(X) and electronegativity E.N(X) are also taken into the account. Of the available various forms of a particular material, care is taken to consider the most stable form by comparing the energies of formation. For a given material, the form having the least energy of formation is considered as stable and considered for analysis.

### 3.1. Features Selection:

The relative importance of various features in classification is found by using Random forests method (RFF) and their relative linear correlation by Pearson correlation heatmap (PCH).

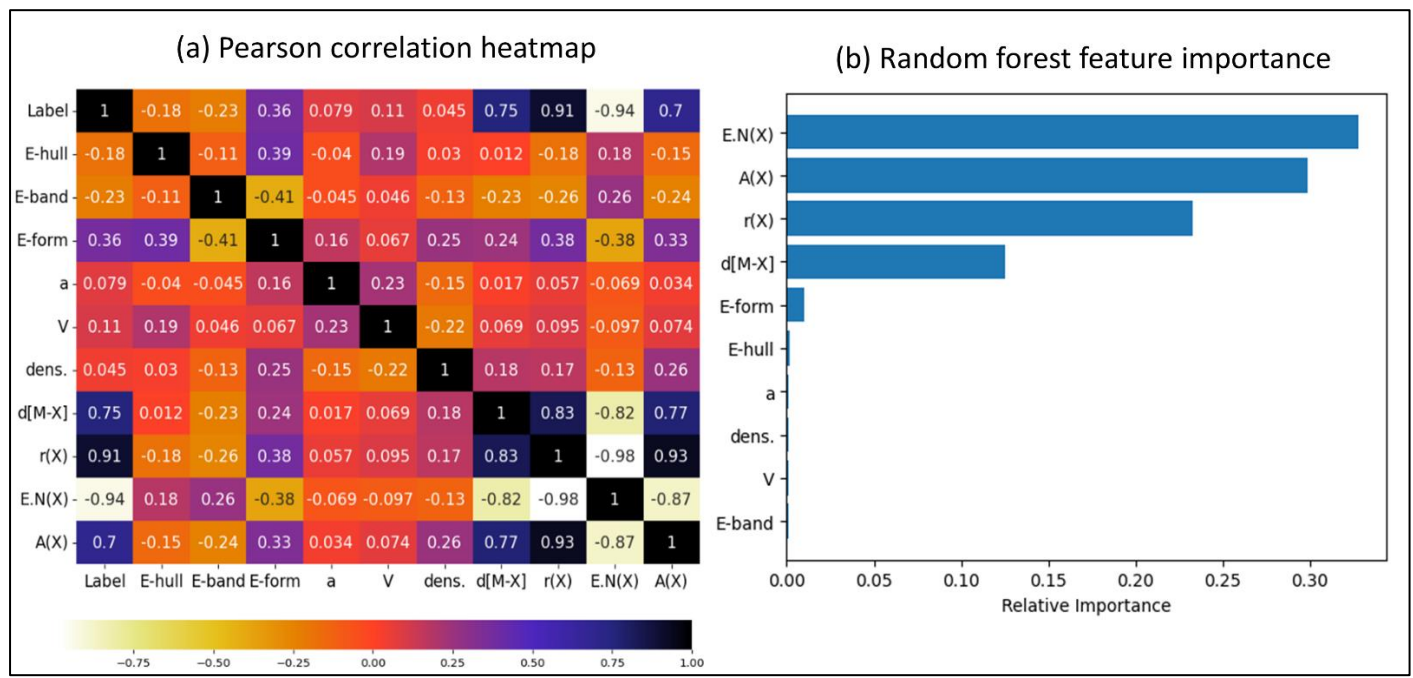

**Figure 2.** (a) Pearson correlation heatmap (PCH) for 350 materials dataset (b) Random forest feature importance (RFF) for 350 materials dataset.

Each Pearson's correlation coefficient (PCC) is defined by:

$$\text{PCC} = \frac{\sum_{i=1}^{n}(x_i-\bar{x})(y_i-\bar{y})}{\sqrt{\sum_{i=1}^{n}(x_i-\bar{x})^2}\sqrt{(y_i-\bar{y})^2}} \tag{10}$$

where $x_i, y_i$ denote the values of two input features and $\bar{x}, \bar{y}$ are their mean values. If the correlation is greater than 0.8, only one among the two features is selected and the other is excluded. The Random forest feature importance and PCH heat map of each feature for 350 materials data set is shown in the Figure 2. For the other two datasets i.e., 102 materials and 192 materials, the figures corresponding to PCH and RFF are given in the supporting information as Figure S1 and Figure S2.

The top 5 features emerging from Random forest feature importance for 350 materials are E.N(X), A(X), r(X), d(M-X) and (E-form). From Pearson Correlation heatmap we observe that the linear correlation among [E.N(X),A(X)], [E.N(X),r(X)] and [At.No.(X),r(X)] is very high (>0.9). Therefore, the effective way is to consider E.N(X) and exclude the other two features. Finally we have considered three features for constructing the ML model: i) electronegativity of the chalcogen atom in the material 'E.N(X)'; ii) distance between transistion metal and chalcogen atom of a given material 'd[M-X]' and iii) energy of formation of a given material 'E-form'. Further analysis has been done using these features.

## 3.2. Classical Machine Learning:

There are a number of Classical ML algorithms available for classification purposes.[60] The available algorithms are very efficient and are applied for a number of tasks in material science. We test our three datasets on 'Support Vector Classifier (SVC)' which is an effective Classical ML model specifically used for small datasets. SVC seeks to find the optimal hyperplane in the N dimensional feature space that seperates the datapoints into different classes.[61] The datasets are then trained on SVC imported from Scikit Learn[62] where 80% is used for training the model and 20% is used for testing the model.

The Classical ML models achieves 100% accuracy in all the three cases showing the feasibility and accuracy of the existing models. In order to address the over-fitting in these models we further performed k-fold cross validation [63] where the entire data is split into k number of subsets or folds ( k $\geq$ 10) and then the model is trained on (k-1) subsets and tested on the remaining one. This is repeated k times to cover all the datapoints. This method provides the more robust estimate of the model's performance since every observation is used for testing and training. The scores achieve 99% accuracy in all the three cases asserting the results from previous SVC approach and negating the doubt of over-fitting.

### 3.3. Quantum Machine Learning:

The problem at hand is to test the performance or ability of the existing QML algorithms viz., VQC, QSVC and QNN with different feature maps on three datsets of different sizes to classify a given material as TMCs or TMOs. In this study, we have used four various feature maps namely Z feature map, ZZ feature map , Pauli ['Z','XX'] and Pauli ['Z','XY'].The most simple of the four without any entanglement among the qubits is Z feature map. It is shown in the Figure 3.

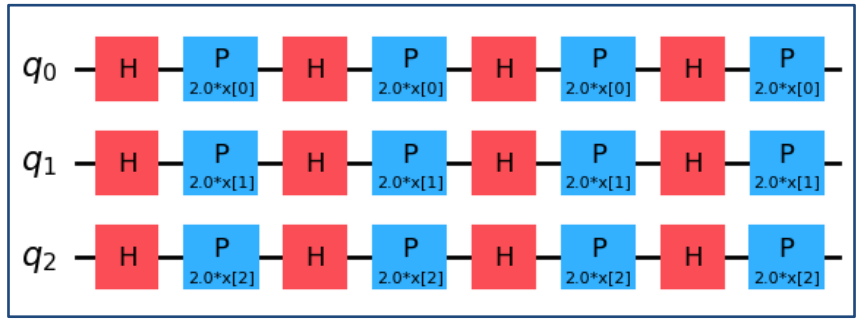

**Figure 3.** Z Encoding Feature Map.

In each QML model, the entire data is split into two parts with 80% used for training the model and 20% is for testing.The feature maps and ansatzes are constructed and the objective function value vs. iteration graph is obtained by optimizing the model performance. In all the cases,we have used 3 qubits to construct the quantum circuits representing the 3 features selected intially. Every circuit is parameterized with the normalized classical features. Figure 4 shows the different feature maps having various degrees of entanglement with 3 qubits used in this study.

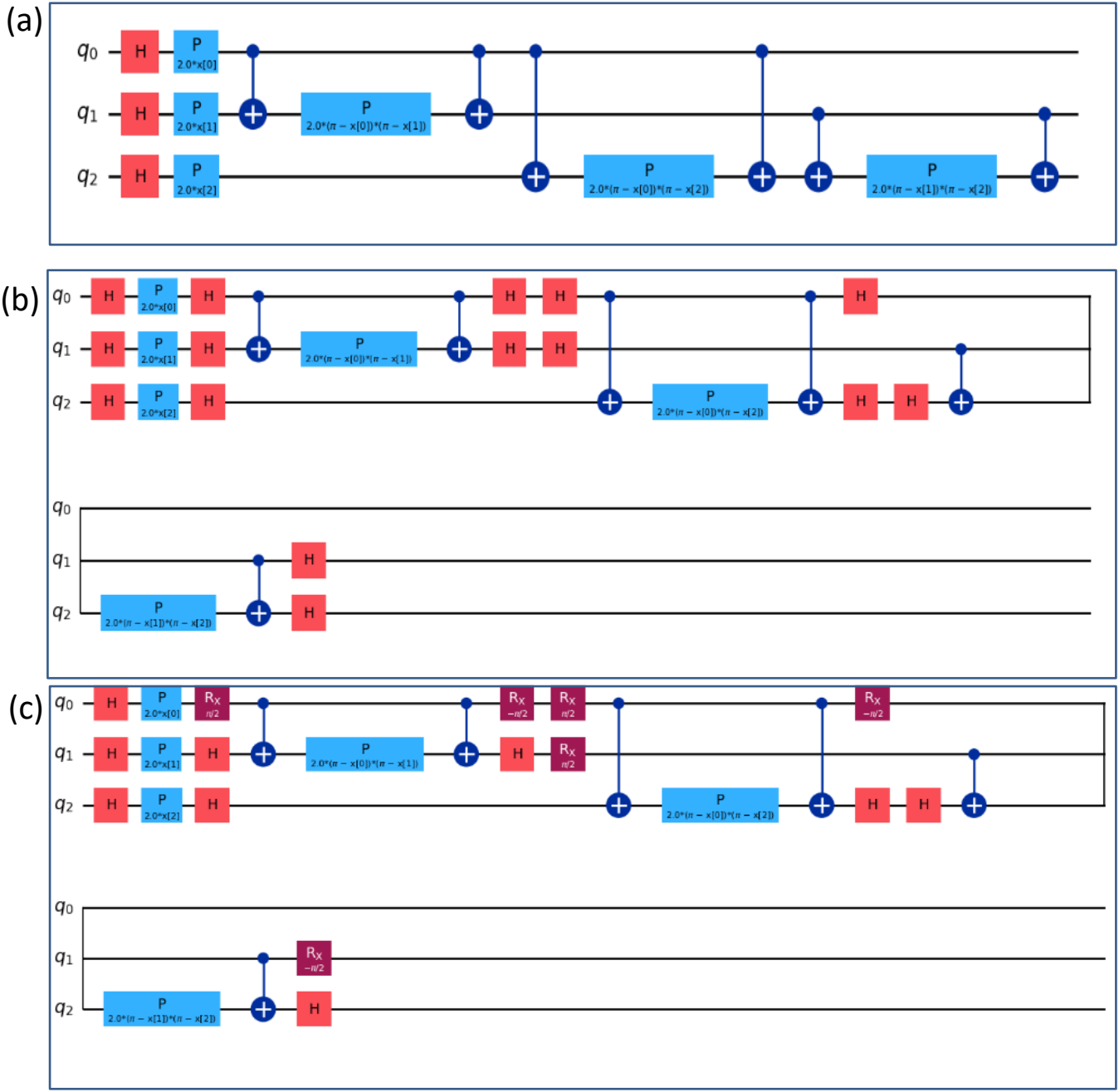

**Figure 4.** (a) ZZ Encoding feature map (b) Pauli [Z,XX] encoding feature map (c) Pauli [Z,XY] encoding feature map. All the circuits have 3 qubits representing the 3 features used in our models.

For training the quantum machine learning models on various classifiers, we have considered the real amplitude ansatz with a circuit depth of 3-layers. Figure 5 shows the real amplitude ansatz used in this study for training VQC and QNN classifiers.

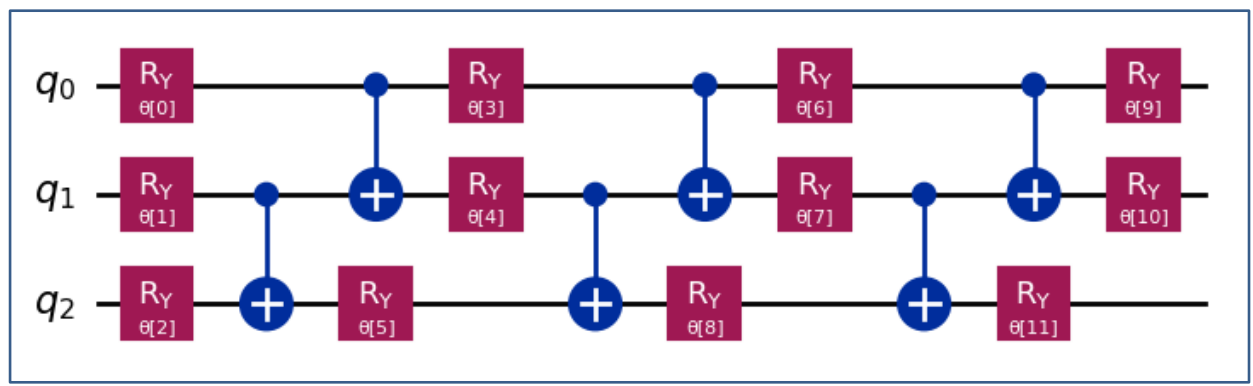

**Figure 5.** Real amplitude ansatz with a circuit depth of 3 layers used in this study for VQC and QNN.

The callback graphs obtained from VQC and QNN for 350 materials dataset are presented in Figure 5.Each possible combination is set for 200 iterations for minimizing the cost function after which the test and train scores of quantum machine learning models are obtained. In all the optimization processes, COBYLA optimizer is used. The most optimal results are obatined with ZZ-Feature map encoding. Table 1 shows the results for ZZ Feature map encoding tested on all the classifers with different sizes of datasets.

By comparing the classification scores of CML model and QML models,we can conclude that the existing QML models can efficiently classify the given material datsets with high accuray.In few cases, the results attain 100% accuracy implying that the QML models are competitive enough for classification of the materials into specific categories given their measured properties.The performance of the model can be improved by increasing the number of datasets and hypertuning their parameters.

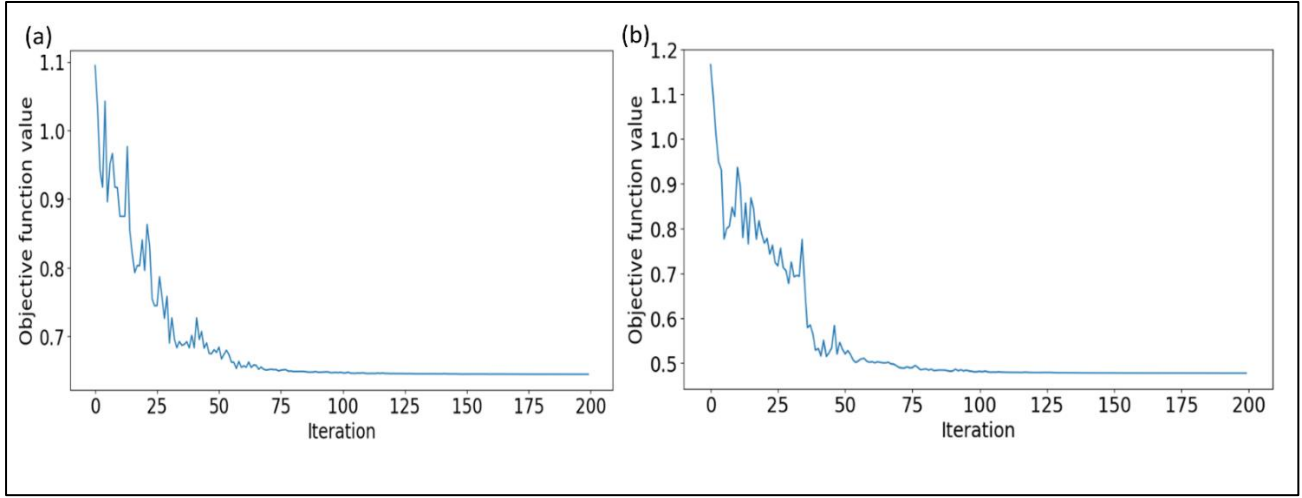

**Figure 5.** (a) Call back graph obtained in VQC with 200 iterations. (b) Call back graph obtained in QNN with 200 iterations.

The comparision of the performance of various QML classifiers leads us to that QSVC which employs kernel estimation in classifying the data gives best results followed by QNN and VQC.This trend can be observed in the case of other feature maps also apart from the widely used ZZ feature encoding.The possible reason being the direct use of the large hyperspace obtained from mapping the data into Hilbert space of qubits is employed for classifying the datapoints using support vectors without again training them on ansatz.This model achieves accuracy of 99% and 98% for test and train datasets, respectively for 350 materials dataset. But, we observe that VQC has the most reliable scores of 93% and 91% when encoded with ZZ feature mapping on 350 materials datset. Thus, for all practical quantum chemistry and materials problems, VQC is employed even though the datset is limited and small.

**Table 1.** The train and test score percentage is tabulated for three different dataset sizes viz., 102 materials, 192 materials and 300 materials when performed on three QML algorithms viz., VQC, QSVC, QNN. For all the cases 'ZZ Feature map encoding' is used.

| Quantum Algorithm/ Classifier | Dataset size | Train score percentage | Test score Percentage |
|---|---|---|---|
| VQC | 102 | 84 | 100 |
| | 192 | 89 | 90 |
| | 350 | 93 | 91 |
| QSVC | 102 | 100 | 100 |
| | 192 | 100 | 97 |
| | 350 | 99 | 98 |
| QNN | 102 | 98 | 95 |
| | 192 | 94 | 95 |
| | 350 | 93 | 94 |

We also note the existence of over-fitting inside few models. Currently, the issue of addressing the over-fitting in QML models is in developmental stage and difficult to employ for practical purposes.[64] We observe that the performance is 100% in the cases where entaglement is least or heavy confirming the presence of over fitting of the model. This can be seen in Table S1, Table

S2 and Table S3 presented in the supporting information. When Z feature map is used, the entanglement is zero and for Pauli[Z, XX] and Pauli[Z,XY] feature maps the entaglement is heavy and hence over fitting is assumed to be influencing the test and train performance for achieving 100% accuracy in many combinations.[65,66] The accuracy of each such combination for classification of materials also depend on the expressibility and entanglement capability of the feature maps as well as the ability of a particular QML classifier to classify the given data.

Further, QML itself is a newly evolving field where researches are trying to extend its utility over a wide variety of problems.Particulary in materials research, this can revive Density functional theory or such areas if efforts are made to inculcate atomistic factors such as interaction forces, potentials and quantum effects.The modelling of materials in various scales becomes more accuarte and easier with the new methodolgies coming into picture in this feild.We also need to identify and explore the ways how QML can represent the unique properties of each material and actively predict their abilities. QML can be an irreplacable asset especially when dealing with the quantum data and its classification where the classical computers fail to tackle. Further developments in creating fault tolerant quantum computers and quantum error mitigation techniques can potentially make QML a powerful tool for future research in science and technology.

## 4. Conclusions:

In summary, we have analysed a hybrid classical-QML approach utilizing VQC, QSVC and QNN algorithms for classifying TMCs and TMOs of the type $M_a X_b$ (a, b = 1,2,3,4; M = transitional metal and X= O, S, Se, Te). Our models were trained on 3 different sizes of datasets (102,192 and 350 materials) where we have labelled TMCs as +1 and TMOs as -1. Through random forests feature importance and Pearson's correlation heatmap, we have identified the three most important features from the initial 10 features and tested them on classical SVC to obtain 100% accuracy. We verified the performnace with k-nearest neighbours method to address overfitting.We utilised these three features to construct the feature maps and ansatz to test on different classifiers.We obtained the best performing combination in the case of ZZ Feature encoding with QSVC achieving 99% and 98% accuracy on test and train data, respectively.These results highlight the potential of QML models in active materials classification which conclusively proves the ability of these models in classification tasks.This work can be further extended to verify on actual quantum computer.These results can give a direction for employing the QML in materials research where the classical ML models fail to investigate the complex chemical space.The intriguing aspect is the issue of over-fitting in QML models which needs a careful endeavour to address and further development needs to be made to identify and resolve .The present study proves that QML can compete if not overtake classical ML in near future with potential applications to classification, regression and related taks for handling huge datsets with complex nature to achieve optimal results within a short computational time period.

**Supporting Information**

Supporting Information is available online or from the author.

**Conflict of Interest**

The authors declare no conflict of interest.

**Data Availability Statement**

The data used in this study is available from the corresponding author upon reasonable request.